\documentclass[preprint2]{aastex6}

\newcommand{\Rmnum}[1]{\expandafter\@slowromancap\romannumeral #1@}
\usepackage{graphicx,epsfig,fancyhdr,psfig,rotating,amsmath,epsf,txfonts,natbib,epstopdf,multirow}
\def \kms {{\rm km\;s$^{-1}$}}

\def \arcsec {$^{''}$}
\def \siiv {Si\,{\sc iv}}
\def \cii {C\,{\sc ii}}

\def \mgii {Mg\,{\sc ii}}

\def \chisq {$\chi^2$}

\def\paper1{{ Paper I}}
\graphicspath{{./figs/}} 

\accepted{for publication in ApJ}

\begin{document}
\title{Transition region loops in the very late phase of flux-emergence in {\it IRIS} sit-and-stare observations}
\author{
Zhenghua Huang, Bo Li, Lidong Xia, MiJie Shi, Hui Fu, Zhenyong Hou
}
\affiliation{Shandong Provincial Key Laboratory of Optical Astronomy and Solar-Terrestrial Environment, Institute of Space Sciences,
Shandong University, Weihai, 264209 Shandong, China; {\it z.huang@sdu.edu.cn}}

\begin{abstract}
Loops are one of the fundamental structures that trace the geometry of the magnetic field in the solar atmosphere.
Their evolution and dynamics provide a crucial proxy for studying how the magnetized structures are formed and heated in the solar atmosphere.
Here, we report on spectroscopic observations of a set of transition region loops taken by the Interface Region Imaging Spectrograph (IRIS) at \siiv\ 1394\,\AA\ with a sit-and-stare mode.
The loops are corresponding to the flux emergence at its very late phase when the emerged magentic features in the photosphere have fully developed.
We find the transition region loops are still expanding and moving upward with a velocity of a few kilometers per second ($\lesssim$10\,\kms) at this stage.
The expansion of the loops leads to interactions between themselves and the ambient field, which can drive magnetic reconnection evidenced by multiple intense brightenings, including transition region explosive events and IRIS bombs in the footpoint region associated with the moving polarity.
A set of quasi-periodic brightenings with a period of about 130\,s is found at the loop apex, from which the \siiv\ 1394\,\AA\ profiles are significantly non-Gaussian with enhancements at both blue and red wings at Doppler velocities of about 50\,\kms.
We suggest that the transition region loops in the very late phase of flux emergence can be powered by heating events generated by the interactions between the expanding loops and the ambient fields and also by (quasi-)periodic processes, such as oscillation-modulated braiding reconnection.
\end{abstract}
\keywords{Sun:atmosphere --- Sun: transition region --- Sun: corona --- techniques: spectroscopic --- Sun: magnetic fields}


\section{Introduction}
\label{sect_intro}
The solar atmosphere is occupied by magnetized structures owing to the coupling of the ionized plasma and the solar magnetic field.
Therefore, the processes of flux emergence are crucial to understand the creations of the magnetized structures in the solar atmosphere.
As one of the prominent structures of the solar atmosphere, loops, are one important tracers of the geometry of the magnetic field.
Consequently, the dynamics of loops provides a proxy for studying the magnetic activities including the processes of flux emergence and heating of plasma that confined in the flux tubes\,\citep[see e.g.][and references therein]{2014LRSP...11....4R}.

\par
At the early stage of flux emergence, the associated flux tubes can appear as serpentine or U-shaped geometries\,\citep{2001ApJ...554L.111F,2004ApJ...614.1099P,2007A&A...467..703C,2008ApJ...687.1373C,2009ApJ...691.1276A,2009ApJ...701.1911P,2014LRSP...11....3C,2018ApJ...853L..26H}
and thus are accompanied by a variety of small-scale energetic events such as Ellerman bombs\,\citep[e.g.][]{1917ApJ....46..298E,2013JPhCS.440a2007R,2013ApJ...779..125N,2015ApJ...798...19N} and UV bursts\,\citep[e.g.][]{2014Sci...346C.315P,2015ApJ...812...11V,2016ApJ...824...96T,2017ApJ...836...52Z,2019ApJ...875L..30C}, 
which are signatures of magnetic reconnection in the lower solar atmosphere\,\citep[see e.g.][]{2006ApJ...643.1325F,2009ApJ...701.1911P,2013ApJ...779..125N,2014Sci...346C.315P,2016MNRAS.463.2190N,2017ApJ...836...52Z,2018ApJ...852...95N,2018PhPl...25d2903N}. More details of small-scale dynamics in the solar transition region associated with flux emergence can be found in a few recent reviews\,\citep{2017RAA....17..110T,2018SSRv..214..120Y,2019STP.....5b..58H}.
Therefore, the evolution and dynamics of loops in the lower solar atmosphere are crucial for understanding how the magnetic flux is emerged and how the confined plasmas are heated in the early ages of the loops.

\par
In a previous work\,\citep[][hereafter \paper1]{2018ApJ...869..175H}, we analysed observations of a loop system above a region with flux emergence at late phase taken by the Interface Region Imaging Spectragraph\,\citep[IRIS,][]{2014SoPh..tmp...25D}, Hinode\,\citep{2007SoPh..243....3K} and the Solar Dynamics Observatory\,\citep[SDO,][]{2012SoPh..275....3P}.
Here, we summarize the major properties of the loop system obtained in \paper1\ that are necessary in order to understand the observations shown in the present work.
The loop system consists of dynamic threads forming at different temperatures from tens of thousand to more than a million Kelvin.
The transition region counterpart of the loop system show a systematic blue-shifts of about 10\,\kms\ near its apex, while about 10\,\kms\ red-shifts are shown in its coronal counterpart.
The electron densities are found to be about $5\times10^{10}$\,cm$^{-3}$ in the transition region loop thread and $1-4\times10^9$\,cm$^{-3}$ in its coronal counterparts.
The north footpoint of the loop system links to a negative polarity that emerged at the edge of and moved away from the main positive one.
The observations also reveal that heating is concentrated in the footpoints of the loop system suggested by their strong coronal emission.

\par
In \paper1, the global view of the loop system has been analysed with the IRIS spectral data taken in the raster mode.
In the present work, we focus on the IRIS sit-and-stare observations taken after the raster one.
During the observing period, the flux emergence in the photosphere as seen by HMI has reached its very late phase as the major emerged magnetic features in the photosphere have fully developed (see Figure 2 and the associated animation of \paper1\ shown observations between 19:38\,UT and 22:58\,UT).
These observations provide temporal evolution of a particular location of the loop system viewed in specified spectral lines.
These observations provide different point of view to the loop system and then
help understand the heating processes of the loop system in the very late phase of flux emergence.
In what follows, we will give the description of the data in Section\,\ref{sect:obs}, the results and discussion in Section\,\ref{sect:res} and the conclusion in Section\,\ref{sect:conclusion}.

\begin{figure}
\includegraphics[clip,trim=0.8cm 1cm 2cm 1cm,width=\linewidth]{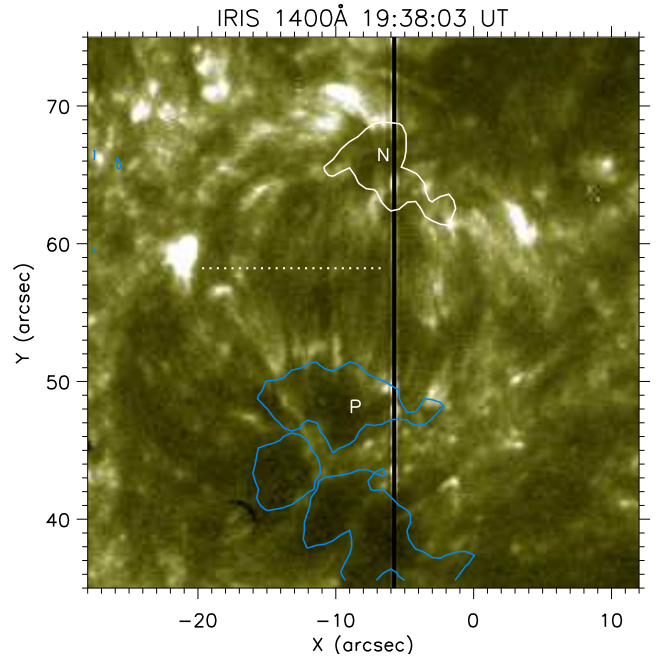}
\caption{The IRIS SJ 1400\,\AA\ image of the loop system. The location of the spectrograph slit is shown as the black vertical line,
where the spectral data were achieved. 
The blue contour curves mark the magnetic features with flux density at level of $800$\,Mx\,cm$^{-2}$, and the white contour curves mark the magnetic features with flux density at level of $-600$\,Mx\,cm$^{-2}$.  
The loop system discussed in the text is connecting two polarities with opposite sign of magnetic flux that are denoted by ``P'' and ``N''.
The white dotted line mark the position of the slice that used to produce the space-time plot in Figure\,\ref{fig:si13ppm}b.
}
\label{fig:slitpos}
\end{figure}

\section{Observations and data analysis}
\label{sect:obs}
The data analysed here were taken by IRIS on 2017 August 26 from 19:38\,UT to 22:58\,UT.
IRIS obtained both slit-jaw (SJ) images and spectral data in sit-and-stare mode.
The compensation to the solar rotation has been applied and thus the spectrograph slit was tracking the same location of the sun when taking the observations.
The SJ images have a pixel size of 0.17\arcsec$\times$0.17\arcsec\ and a cadence of 49\,s.
The spectral data were taken with a 0.33\arcsec\ wide slit, an exposure time of 15\,s and a step cadence of about 16.4\,s.
The pixel size along the slit of the spectral data is 0.17\arcsec.
We analyse the IRIS level 2 data that have been calibrated by the instrument team.
In addition, we apply the despike procedure,  {\it iris\_prep\_despike.pro} in the IRIS data analysis package of {\it solarsoft}, to remove the spike pixels affected by cosmic rays.
We have manually checked through the image and confirmed that this procedure did not remove any good data points in the observations.
We also convert the unit of the spectral data from DN to photons, for which we use the gain factor of 4\,photons\,DN$^{-1}$ as given by \citet{2018SoPh..293..149W}.

\par
The loop system studied in this work can be seen in images taken by the IRIS SJ 1400\,\AA\ passband as shown in Figure\,\ref{fig:slitpos}.
It contains multiple loops with length of $\sim$15\arcsec, and connects the two polarities with opposite sign of magnetic fluxes as revealed by the line-of-sight magnetograms measured by the Helioseismic and Magnetic Imager\,\citep[HMI,][]{2012SoPh..275..207S} aboard SDO (see Figure\,\ref{fig:slitpos}).
The spectrograph slit extends in the solar south-north direction and it was almost (but not strictly) aligned with the threads of the loop system.
The spectral data have 778 pixels along the slit corresponding to 130\arcsec\ in the solar south-north direction and 735 time steps that is about 3.4 hours.
The spectral data include eight spectral windows that include major spectral lines covered by IRIS, such as \mgii\ h\&k, \cii\ 1335\& 1336\,\AA\ and \siiv\ 1394\& 1403\,\AA.
In this study, we mainly focus on the transition region counterparts of the loop system that are observed in the \siiv\ 1394\,\AA\ (corresponding to a temperature of $\sim7.9\times10^4$\,K).

\par
\begin{figure}
\includegraphics[clip,trim=1.6cm 1.5cm 0.8cm -0.5cm,width=\linewidth]{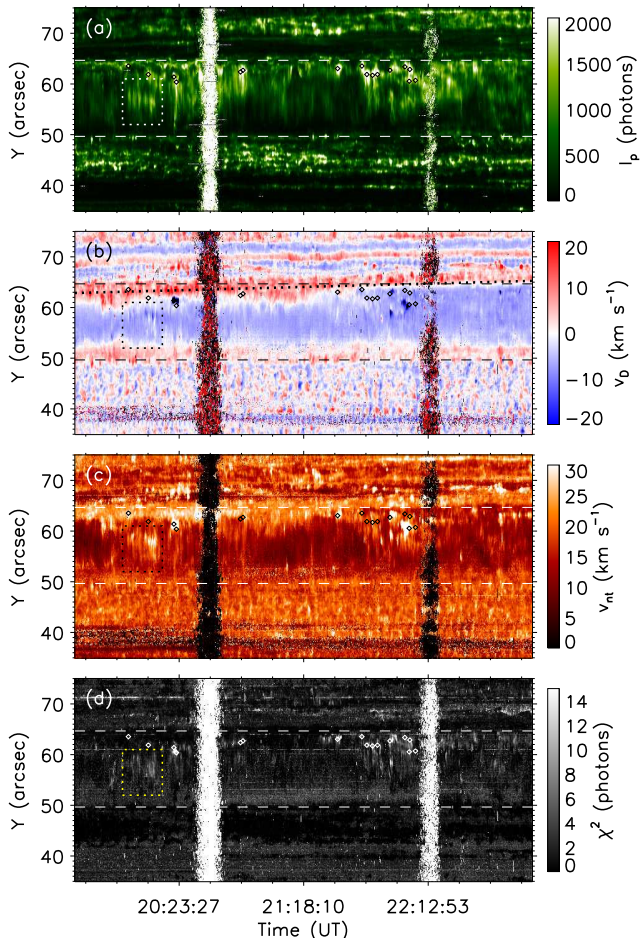}
\caption{Space-time maps of peak intensities (a), Doppler velocities (b), Nonthermal velocities (c) and reduced $\chi^2$ (d)  derived from \siiv\ 1394\,\AA\ emission of the region of and around the loop system.
The interval between two neighbouring minor ticks in time axis is 13 minutes and 41 seconds.
The dashed lines mark the locations that roughly correspond to the footpoints of the loop system.
The diamond symbols denote the locations of the compact brightenings in \siiv\ intensity map.
The thick dotted line in panel b mark the linear fit to the gravitational centers of the redshifted patterns (see the main text for details).
The region enclosed by thin dotted lines is further investigated in Section\,\ref{subsec4}.
The two data gaps around 20:36\,UT and 22:13\,UT were caused by influence of energetic particles
while the spacecraft passing the South Atlantic Anomaly.
}
\label{fig:si13sp}
\end{figure}

To derive Doppler velocities and non-thermal velocities, we follow similar routines as described in \paper1.
Single Gaussian fits are applied to the spectra of all the pixels, which return peak intensity, line centre and line width.
The rest wavelengths were obtained from the average spectral profiles of pixels from 50 to 700 along the slit (almost the entire slit) and time steps from 250 to 520 (between the two data gaps due to the South Atlantic Anomaly as displayed in Figure\,\ref{fig:si13sp}, i.e. one IRIS orbit).
While fitting the data with Gaussian fit function in IDL (v7.0) standard library, we also obtained the reduced chi-square ($\chi^2$) statistic that represents the goodness-of-fit and can be expressed as
$$\chi^2=\frac{1}{N-f}\sum_{i=1}^N\frac{(D_i-G_i)^2}{G_i},$$
where $N$ is the number of data points for fitting, $f$ is the number of fitting coefficients, $D_i$ is the value of the data point indexing of $i$ given in the observations, and $G_i$ is the value of the data point indexing of $i$ given by the Gaussian fits.
In many cases, the $\chi^2$ evaluates the statistical significance of fit error rather than tests the fitting models.
Therefore, here the $\chi^2$ is used only as a guide for searching non-Gaussian profiles and those are then selected manually.

\begin{figure}
\includegraphics[clip,trim=0.5cm 2cm 0cm 0cm,width=\linewidth]{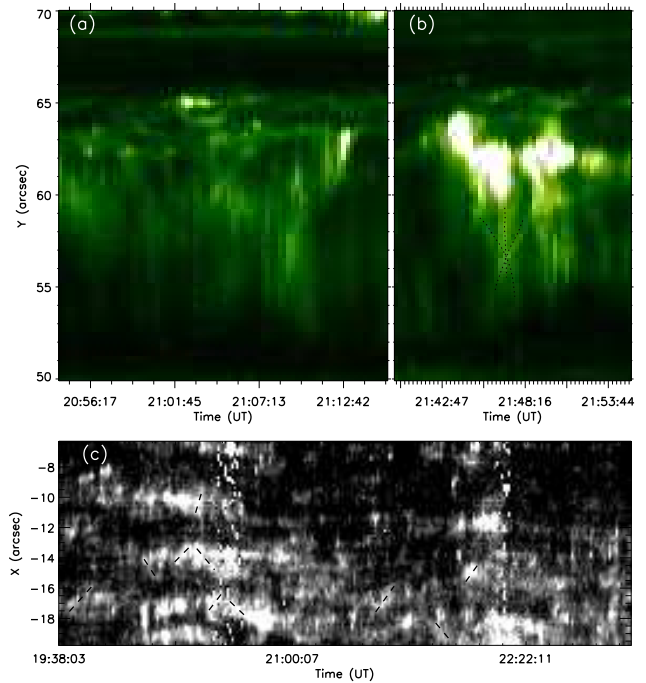}
\caption{Panel (a)-(b): close-up views of two subregions of Figure\,\ref{fig:si13sp}a.
The black dotted lines in panel (b) present an example of three crossing patterns in the space-time map.
Panel (c): the space-time plot of the SJ 1400\,\AA\ images for the slice shown in Figure\,\ref{fig:slitpos}.
We randomly trace a few transverse motions of the loops and they are marked by the black dashed lines in panel (c).}
\label{fig:si13ppm}
\end{figure}

\section{Results and interpretations}
\label{sect:res}
In Figure\,\ref{fig:si13sp}, we show intensity, Doppler velocity, nonthermal velocity and \chisq\ maps derived from the \siiv\ 1394\,\AA\ sit-and-stare observations.
Those derived from \siiv\ 1403\,\AA\ also show identical features as in Figure\,\ref{fig:si13sp},
and we do not show them in this paper.

\par
From the intensity map (Figure\,\ref{fig:si13sp}a), we can see that bright threads appear repetitively,
suggesting a dynamic nature of the transition region loops.
The brightness along the loops has a systematic bias toward the north footpoint.
Compact brightenings can be seen and they are all found to locate in the places close to the north footpoint (see those denoted by diamond symbols in Figure\,\ref{fig:si13sp}).
Their lifetimes in this observations are less than 2 minutes (most of them are about 1 minute).
These compact brightenings are being investigated and shown in detail later.

\par
In the Doppler velocity map (Figure\,\ref{fig:si13sp}b), we can see clear differences between the main parts and the footpoints of the loops.
In the main parts of the loops including the apexes, persistent blue-shifts with Doppler velocities of $\lesssim$10\,\kms\ are shown.
While both the footpoint regions are red-shifted, the north footpoint region (associated with the negative polarity that moved away from the emerging site at the early stage of the flux emergence) shows red-shifts as large as 20\,\kms,
and the south (associated with the stationary positive polarity with the emerging site occurred at its edges) shows red-shifts of only a few kilometers per second.
The red-shifts in the footpoint regions are fluctuated,
which is most clearly seen in the north footpoint region (see Figure\,\ref{fig:si13sp}b).

\par
The nonthermal velocity map (Figure\,\ref{fig:si13sp}c) shows that on average the region near the north footpoint has larger values than that in the regions near the loop apex and south footpoint.
Consistent with that found in \paper1, the loop region is dominated by small nonthermal velocities ($<$10\,\kms) except the flaring loop threads.
For most of the compact brightenings in the region near the north footpoint, they have much larger nonthermal velocities ($>$30\,\kms).

\par
The $\chi^2$ map (Figure\,\ref{fig:si13sp}d) gives a view of goodness of fits using single Gaussian model.
It is a proxy for searching locations that produce non-Gaussian spectra.
In general, most of the pixels in the loop system can be well fitted by single Gaussian models.
One can see that some of the compact brightenings show large $\chi^2$ values, suggesting their non-Gaussian profiles.

\par
In what follows, we will make further investigation to interpret what we observed.
Based on these analyses, we will give our understandings to the activities related to flux emergence and loop heatings.

\subsection{What we have observed}
\label{subsec1}
The first question to be answered is whether the same loop thread(s) was (were) targeting through out the entire observing period.
In Figure\,\ref{fig:si13ppm}a-b, we show a close-up view of two subregions of Figure\,\ref{fig:si13sp}a.
We can see that most (but not all) of the bright threads in the space-time map of the \siiv\ peak intensity are vertical,
suggesting that the brightness along the slit appeared within the cadence of the sit-and-stare data (i.e. 16.4\,s).
If assuming these bright threads are signatures of propagating plasmas,
they would give velocities more than 600\,\kms\ because some of them extend for about 15\arcsec.
Such plasma speeds seem to be not real in the transition region.
A more plausible scenario is that these bright threads are signatures of active loops moving transversely and passing the field-of-view of the spectral slit.
The transverse motions of the loops with speeds of a few hundred meters per second can be seen in the slit-jaw images and
these motions are either be eastward or westward while the loops are extending in the south-north direction (Figure\,\ref{fig:si13ppm}c).

\par
In Figure\,\ref{fig:si13ppm}b, we can also see crossing of multiple vertical bright features in the space-time map (see those marked by dotted lines). 
This is an evidence that these bright threads belong to different loops that passing the field-of-view of the spectral slit, and
these loops are crossing and/or overlapped each other.
Therefore, in most of the cases we were observing different transition region loops that were actively evolving and moving across the field-of-view of the slit,
although in some cases we also observe moving features in some loop threads.

\subsection{Understanding the Doppler velocities}
\label{subsec2}
The persistent blue-shifts in the main part of the loops indicate that the observed transition region loops were moving upward,
even though the main processes of flux emergence in the photosphere had completed as seen in the magnetograms (see Figure 2 and the associated animation of \paper1).
It demonstrates that the emerged flux tubes need times to expand (and thus moving upward) to have a stable geometry.

\par
The red-shifts in the footpoints could be results of two processes.
One is the downward motion due to the gravity while the loops were raising.
The other one is submergence of small loops that were generated by magnetic reconnection between the emerging loops and pre-existed ones.
We observed that the emerged negative polarity was frequently canceling with opposite polarity while it was moving away from the emerging site (see the associated animation to Figure 2 of \paper1).
As shown in Figure\,\ref{fig:si13sp}, compact brightenings were seen in the north footpoint region, 
these brightenings can be related to the interaction between the loops and the ambient field around the moving polarity.
This process can lead to magnetic reconnection between the emerged loops and small ambient loops creating even smaller loops that will subject to submergence, and thus resulting in stronger red-shifts in the north footpoint region than that in the south.

\par
We can use the redshifted patterns in the north footpoint region to represent the locations of the north footpoints of the loops.
Via applying a linear fit to the gravitational centers of the redshifted patterns at each observing time (see the thick dotted line in Figure\,\ref{fig:si13sp}b), 
we found that the horizontal speed of the north footpoints is 0.14\,\kms.
This represents the separation speed of the two footpoint regions of the loop system.
Assuming an emerging flux tubes is an arc that becomes semicircle at the end,
the separation speed of the two footpoint regions gives the increasing rate of the hypotenuse,
while the Doppler velocity ($\sim$10\,\kms) gives the increasing rate of the height of the loop tops.
Given an emerging semicircle loop with a radius of $r$ and the angle of the arc above the photosphere is $2\theta$, the separation speed of the footpoints will be $v_{horizontal}=2r\frac{d(\sin{\theta})}{dt}=2r\cos{\theta}\frac{d\theta}{dt}$, and the upflow speed of the loop top will be $v_{vertical}=-r\frac{d(\cos{\theta})}{dt}=r\sin{\theta}\frac{d\theta}{dt}$, where $t$ is the time. 
While the observations give $v_{horizontal}\approx 0.14$\,\kms\ and $v_{vertical}\approx 10$\,\kms, it gives $\theta=89.5^\circ$. This indicates that the arc of the loop above the photosphere at this stage has an angle of $179^{\circ}$,  which is consistent with the very late phase of the flux emergence.

\begin{figure*}
\includegraphics[clip,trim=0.5cm 0.5cm 0cm 0cm,width=\linewidth]{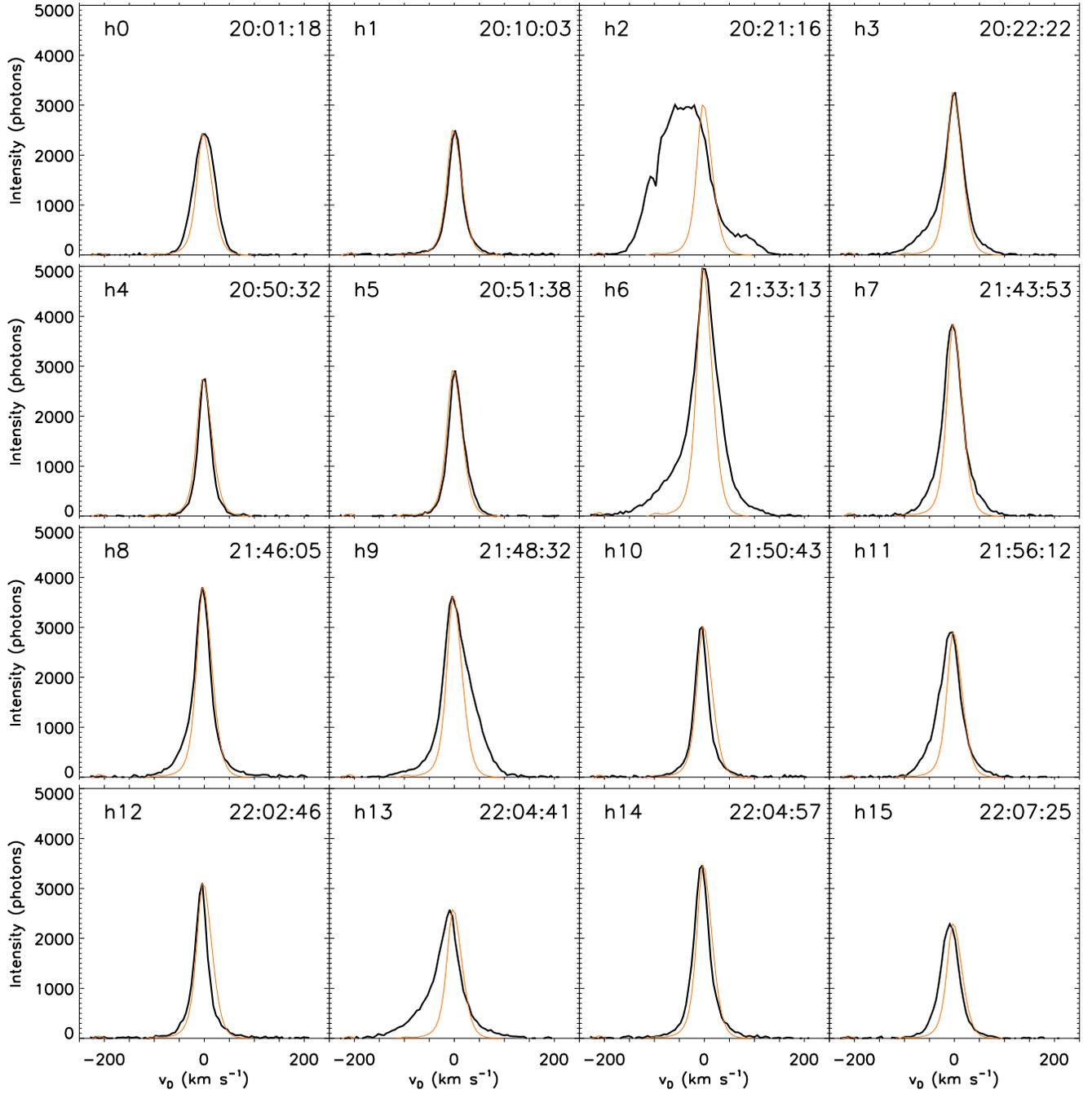}
\caption{Samples of \siiv\ 1394\,\AA\ spectral profiles (black lines) taken from the compact brightenings in the north footpoint region (denoted by diamond symbols in Figure\,\ref{fig:si13sp}).
The order of the samples are sorted by the observing times.
The orange lines are the referent profile (see Section\,\ref{sect:obs}) amplified to have the same peak values as the sample profiles shown in the same panel.
}
\label{fig:si13nfhe}
\end{figure*}

\subsection{Heating events}
\label{subsec3}
As discussed above, the compact brightenings in the north footpoint region could be signatures of heating processes in these loops.
In order to examine what might had occurred in these brightenings, we further exploit their spectral data.
As examples, we show \siiv\ 1394\,\AA\ profiles taken from 16 compact brightenings in Figure\,\ref{fig:si13nfhe}.

\par
In six samples (h3, h6, h7, h9, h11 \& h13), the \siiv\ profiles are non-Gaussian with enhancements in both (either) red and (or) blue wings.
The signatures of non-Gaussian profiles in these events can also be confirmed by the $\chi^2$ map (Figure\,\ref{fig:si13sp}d).
Even though some of these samples show enhancements in both wings, all of them have one wing much stronger than the other.
These \siiv\ profiles are treated as typical spectra that are defined as transition region explosive events\,\citep[see e.g.][]{1991JGR....96.9399D,1997Natur.386..811I,2014ApJ...797...88H,2015ApJ...810...46H,2017MNRAS.464.1753H,2018MNRAS.479.2382L}.
Such profiles can be produced in magnetic reconnection that proceeds via plasmoid instability characterized by multiple magnetic islands and acceleration sites\,\citep{2015ApJ...813...86I}.
Our results could be understood in the similar way of a recent work by 
\citet{2019ApJ...873...79C}, who found that non-Gaussian \siiv\ spectra with enhancement at the blue wing are found in locations on network jets away from the footpoints
and those showing enhancement only at the red wing of the line are often located around the jet footpoints.
Here, the spectral slit might not capture equally magnetic islands moving at both directions and thus the enhancement at two wings are not balance.
The strong emission in the line center is corresponding to heating while the enhanced wings are associated with the moving magnetic islands produced by the plasmoid instability\,\citep{2015ApJ...813...86I}.
Therefore, these six samples are a hint of magnetic reconnection occurred in the north footpoint region of the loop system.
These reconnection processes should take place while the north footpoints of the loops were moving through the ambient fields.
This kind of events have also reported recently by \citet{2018ApJ...856..127G,2019ApJ...871...82G}, who reported a long-lived UV burst in the late phase of flux emergence and believed it was a result of interaction between the emerged and the ambient fields.
Actually, many energetic events in the solar atmosphere can be triggered by the interactions between emerging and pre-existing magnetic fields\,\citep[see][and references therein]{2019arXiv190704296I}.

\par
In one sample (h2), we found a very broad \siiv\ 1394\,\AA\ profile with absorption lines of Ni\,{\sc ii} embedded at its blue wing (see the dip at $v_D$ of about $-100$\,\kms).
Such a \siiv\ 1394\,\AA\ profile owns the same characters of that emitted from a class of UV bursts\,\citep{2018SSRv..214..120Y}, namely hot explosions or IRIS bombs\,\citep{2014Sci...346C.315P,2015ApJ...812...11V,2016ApJ...824...96T}.
IRIS bombs frequently occur in regions with flux emergence at early stage\,\citep{2014Sci...346C.315P,2017ApJ...836...63T,2018ApJ...854..174T}, but are rarely reported in the late phase of flux emergence.
We also notice that the event here is not as intense as that reported in the previous works, suggesting the energy release was not as strong as those at the early stage of flux emergence.
Because we did not find any IRIS bombs in the raster of the region (\paper1) and only one is found in this 3.5 hour sit-and-stare observations,
we can conclude that IRIS bombs are not common in the late phase of this flux emergence.

\par
In nine samples (h0, h1, h4, h5, h8, h10, h12, h14 \& h15), the \siiv\ profiles are similar to the reference, i.e. no signature of non-Gaussian and/or additional broadening.
Because the nonthermal broadening of transition region lines carries signature of heating processes, these nine brightenings seem to be associated with loops that have been just heated rather than that are being heated.

\par
In summary, heating events are not rare in the very late phase of the flux emergence.
Most of them should result from interactions (likely, magnetic reconnection) between the emerged and the ambient field.
Such interaction can produce heating events that are then heat the loops in the transition region.
IRIS bombs might be found but not common in the late phase of this flux emergence, and they are not as intense as those found at the early stage of flux emergence.

\begin{figure*}
\includegraphics[clip,trim=0.5cm 0.5cm 0cm 0cm,width=\linewidth]{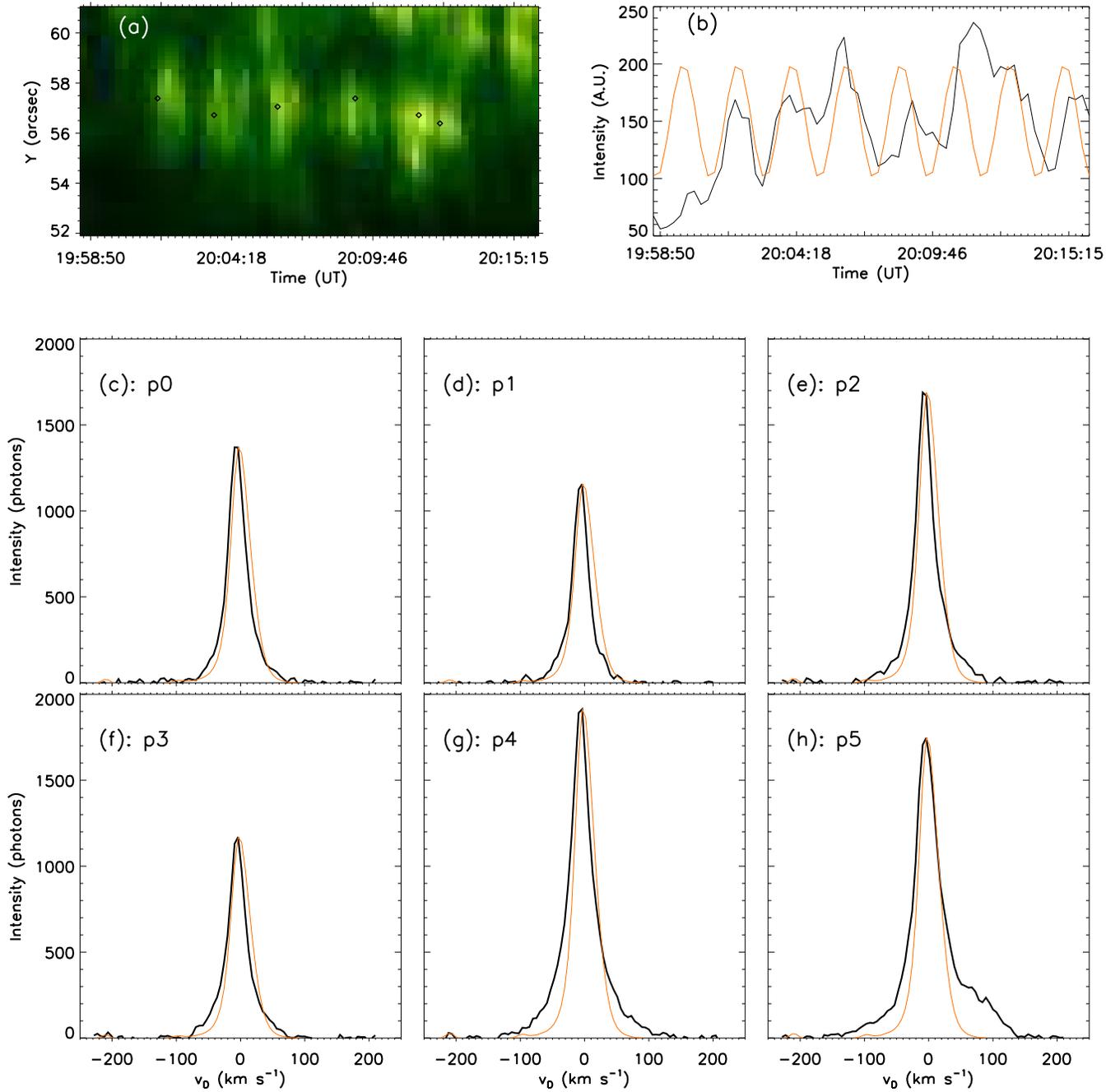}
\caption{A region near loop apex during an observing period when it was showing quasi-periodic variation of brightness.
(a): Space-time map of \siiv\ 1394\,\AA\ intensity, which is zoomed-in view of the region denoted by dotted lines in Figure\,\ref{fig:si13sp}.
The diamond symbols marks the positions, from which the emitted \siiv\ 1394\,\AA\ profiles are given in panels (c--h).
(b):  Temporal variation of the brightness averaged from the spatial range shown in panel (a).
The black line is the observational curve and the orange line is a sine function with a period of 131\,s.
(c--h): Samples of \siiv\ 1394\,\AA\ spectral profiles (black lines) taken from the bright features denoted by diamond symbols in panel (a).
The order of the samples are sorted by the observing times (i.e. from left to right in panel a).
The orange lines are the referent profile amplified to have the same peak value as the sample shown in the same panel.
}
\label{fig:si13qp}
\end{figure*}

\subsection{Quasi-periodic brightenings near the loop apex}
\label{subsec4}
In Figure\,\ref{fig:si13sp}, we can see that the brightness at the loop apex shows a quasi-periodic variation around 20:05\,UT (see the region enclosed by dotted lines).
The length of these brightenings along the slit is about 3\arcsec. 
The nonthermal velocities and $\chi^2$ of the brightenings are relatively large
suggesting non-Gaussian profiles emitted from.
In Figure\,\ref{fig:si13qp}, we show a zoomed-in view of the region, its light curve and samples of \siiv\ 1394\,\AA\ pofiles.
We can see that the brightenings appear only near the loop apex without extending toward the footpoints (Figure\,\ref{fig:si13qp}a).
Moreover, their extension along the slit seems to appear at the same time suggesting that they may not have any apparent propagation along the slit.
The temporal variation of the brightness (Figure\,\ref{fig:si13qp}b) shows a quasi-periodic behavior with a period about 130\,s.
The examples of \siiv\,1394\,\AA\ profiles taken from the brightenings have enhanced wings at Doppler velocities $\gtrsim50$\,\kms\ (Figure\,\ref{fig:si13qp}c--h).
These \siiv\ 1394\,\AA\ profiles have also been found in brightenings in transition region loops\,\citep{2017MNRAS.464.1753H}, which were interpreted as signatures of magnetic reconnection due to braidings.
The enhanced wings are seen in both blue and red wings in all the profiles and in most of the case the enhancements in the two wings of a profile are similar.
It also shows that the profile near the line center is not much different from the referent ones, and thus
the enhanced wings are the main reason why the nonthermal velocities of these features are larger than the reference.
These quasi-periodic brightenings can be evidence of heating mechanisms in this loop system other than the heating events discussed in the previous section.

\par
Based on the periodicity of these brightenings, a question is whether they are standing mode MHD waves.
The first thought is standing sausage waves because they can produce spectral profiles with enhancements at both red and blue wings\,\citep[see e.g.][]{2016ApJ...823L..16T,2019ApJ...870...99S,2019ApJ...874...87S}.
For a loop with a cross-section less than 1\arcsec, the period of sausage mode should be less than 10\,s\,\citep{2003A&A...412L...7N,Nakariakov2005}, and thus the period of the observed phenomena is too large to be a sausage wave.
A possibility is combination of sausage and kink modes, which can occur concurrently in the solar atmosphere\,\citep{2012NatCo...3E1315M}.
The sausage mode can produce the observed spectral profiles, while the kink mode allow the bright loops moving transversely and thus can provide the observed periodicity.
The periodicity of the sausage mode cannot be resolved in the current observations due to the temporal resolution.
A question in this scenario is whether a flux tube consisting of sausage motions with speeds more than 50\,\kms\ is stable.
This should be investigated in the future numerical studies.

\par
The other thought is shocks that can be frequently seen in sunspots having periods of about 3 minutes\,\citep[see][ and references therein]{ 2014ApJ...786..137T,2018ApJ...855...65H}.
Because shocks could be results of leakage of p-modes, they can drive quasi-periodic brightenings in the transition region\,\citep{2015ApJ...804L..27Y,2016A&A...589L...7H,2018ApJ...855...65H}.
While shocks are propagating along loops, they might drive transverse motions (i.e. kink motions) and lead to repetitive appearance and  disappearance of bright loops in the field-of-view of the spectrograph slit.
The observed quasi-periodic brightenings can be responses of shocks interacting with fine structures in the loop apex.
Further theoretical study is required to understand whether and how shocks can produce such phenomena.

\par
Another alternative scenario is that they are produced by repetitive magnetic reconnection.
Such reconnection can result from magnetic braidings due to the plasma motions of the footpoints of the loops in the photosphere\,\citep{1983ApJ...264..635P,1983ApJ...264..642P}.
Because multiple loop threads can be seen in the field-of-view,
it satisfies the basic requirement of this magnetic braiding scenario.
Such magnetic reconnection can produce the observed non-Gaussian profiles of \siiv\,1394\,\AA\,\citep[see e.g.][]{2017MNRAS.464.1753H, 2018ApJ...854...80H}.
Because the driven processes of the magnetic braiding can be modulated by the omnipresent 3-minute oscillations in sunspots,
the braiding-induced magnetic reconnection then occur with a periodicity.
It has been demonstrated by numerical simulations that magnetic reconnection modulated by p-mode waves can produce quasi-periodic transition region explosive events\,\citep{2006SoPh..238..313C}.
Moreover, the period of the brightenings also agree with standing torsional (Alfv\'en) modes in a flux tube with length comparable to the observed loops\,\citep{Nakariakov2005}, and thus, such torsional motions are also candidates of the drivers of the modulation.
The locations of these brightenings suggest that the magnetic braids tend to concentrate at the loop apex. 

\par
As discussed above, the quasi-periodic brightenings can have several interpretations.
A question is which one is the most plausible.
Based on the fact that these brightenings show bi-directional flows with more than 50\,\kms\ speed,
which is comparable to the Alfv\'en speed in these transition region loops (given as about 80\,\kms\ by assuming an electron density of $10^{11}$\,cm$^{-3}$ as measured in \paper1 and a magnetic strength of 10\,G).
This is consistent with the magnetic reconnection theory that can produce bidirectional flows at Alfv\'en speed\,\citep{priest2000book}.
Therefore, the modulated braiding scenario sounds most plausible, although we cannot rule out the rest in the present stage. 

\section{Conclusion}
\label{sect:conclusion}
In the present study, we analysed spectroscopic observations of a transition region loop system taken with IRIS sit-and-stare mode in \siiv\ ($\sim7.9\times10^4$\,K).
The loop system contains loops with length of $\sim$15\arcsec.
It was corresponding to flux emergence in the very late phase that the emerged magnetic features in the photosphere have fully developed as seen in the HMI magnetograms.
During the observing period, the distance between the two polarities of the loop system was increasing, which the polarity associated with the north footpoint was moving away from the more stationary south one.
Based on these observations, we analysed the intensities, Doppler velocities, nonthermal velocities and goodness of fits ($\chi^2$) of the region seen in the \siiv\ spectral data. 
We aim to study the dynamics in this loop system, which can help understand the heating processes in the transition region loop system at very late phase of flux emergence.

\par
The sit-and-stare observations show repetitive bright threads, suggesting a dynamic nature of the loop system.
In the intensity map, most of the bright threads appear to be vertical, and we found they were most likely bright loops moving transversely and passing through the slit.
Frequently, we can also observe that multiple loops are passing the same location at the same time and they contribute to one bright feature in the observations.

\par
The Doppler velocities show a persistent blue-shifts of $\lesssim$10\,\kms\ in the main part of the loops including the apexes.
This result indicates that the emerged loops are still moving upward, which is very likely caused by the expansion of the loops while their footpoints are moving away from each other.
The upward moving loops can also result in plasma in the loops moving toward the footpoints and thus present as red-shifts in the footpoints region as observed.
The red-shifts in the north footpoint region of the loops are significantly larger than that in the south.
While the north footpoint region is associated with the moving polarity, we suggest that the extra red-shifts are results of submergence of smaller loops produced by magnetic reconnection between the loops and the ambient fields.

\par
The magnetic reconnection between the loops and the ambient fields is also evidenced by abundant intense brightenings found in the north footpoint region.
The \siiv\ 1394\,\AA\ profiles of some brightenings show a clear signature of non-Gaussian with enhancements in both (either) red
and (or) blue wings, and 
we suggest that they are evidence of occurrence of magnetic reconnection.
The \siiv\ 1394\,\AA\ profiles of some other brightenings are Gaussian without additional broadening, which is the responses of post-heated loops rather than the heating processes themselves.
We observed one brightening has characteristics of IRIS bombs in the \siiv\ 1394\,\AA\ profile.
It demonstrates that IRIS bomb can also (although rare) occur at the very late phase of flux emergence.
The IRIS bomb occurred at the very late phase of flux emergence is not as intense as those at early phase as reported previously.
We suggest that most of these brightenings are heating events in the transition region loops resulting from interactions between the emerged and the ambient field.

\par
During a part of observing period, we found an example of quasi-periodic brightenings (with a period of $\sim$130\,s) at the loop apex, which also show larger nonthermal velocities and $\chi^2$ values.
The \siiv\ 1394\,\AA\ spectral profiles from the quasi-periodic brightenings are non-Gaussian with weak enhancements in both the red and blue wings at Doppler velocities $\gtrsim50$\,\kms.
The enhancements at the red and blue wings are almost identical in most of the cases.
Such a quasi-periodic brightenings might evince some other dynamic processes that can contribute to heating in the loop system.
We interpret this quasi-periodic brightenings as combination of sausage and kink standing mode, repetitive shocks that associated with 3-minute oscillations or braiding-induced magnetic reconnection that modulated by the 3-minute oscillations.

\par
In summary, the loop system corresponding to the flux emergence in the very late phase is still expanding and moving upward in the transition region.
The expansion of the loop system leads to interactions between the loops and the ambient fields that can create the dynamics in the loops including heating events in the transition region.
Meanwhile, (quasi-)periodic processes, such as MHD oscillations, shocks and oscillation-modulated braiding reconnection, might also contribute to heatings in the loop system.

\acknowledgments
{\it Acknowledgments:}
We would like to thank the anonymous referee for the constructive comments.
This research is supported by National Natural Science Foundation of China (U1831112, 41627806, 41604147, 41674172, 11761141002),
and the Young Scholar Program of Shandong University, Weihai (2017WHWLJH07).
The observation program at BBSO was supported by the Strategic Priority Research Program --
The Emergence of Cosmological Structures of the Chinese Academy of Sciences, Grant No.
XDB09000000.
Z.H. is grateful to BBSO, IRIS and Hinode operating teams for their help and to the BBSO staff for their hospitality while carrying out the observing campaign.
IRIS is a NASA small explorer mission developed and operated by LMSAL with mission operations executed at NASA Ames Research center and major contributions to downlink communications funded by ESA and the Norwegian Space Centre.
Courtesy of NASA/SDO, the AIA and HMI teams and JSOC.
CHIANTI is a collaborative project involving George Mason University, the University of Michigan (USA) and the University of Cambridge (UK).

\bibliographystyle{aasjournal}
\bibliography{bibliography}

\end{document}